\documentclass[aps,pra,twocolumn,superscriptaddress,showpacs,a4paper]{revtex4}

\usepackage{graphicx,amsmath,amssymb}

\begin{document}

\title{A random laser tailored by directional stimulated emission}

\author{M. Leonetti}
\affiliation{Instituto de Ciencia de Materiales de Madrid (CSIC), Cantoblanco 28049 Madrid Espa\~{n}a.}

\author{C. Conti}
\affiliation{Research Center INFM-CNR, c/o Universit\'{a} di Roma Sapienza, I-00185, Roma Italy}

\author{C. L\'{o}pez}
\affiliation{Instituto de Ciencia de Materiales de Madrid (CSIC), Cantoblanco 28049 Madrid Espa\~{n}a.}
\email{marco.leonetti@icmm.csic.es}
\date{\today}

\begin{abstract}
A disordered structure embedding an active gain material and able to lase is called random laser (RL). The RL spectrum may appear either like a set of sharp resonances or like a smooth line superimposed to the fluorescence. A recent letter \cite{Leonetti2011} accounts for this duality with the onset of a mode locked regime in which increasing the number of activated modes results in an increased inter mode correlation and a pulse shortening ascribed to a synchronization phenomenon. An extended discussion of our experimental approach together with an original study of the spatial properties of the RL is reported here.
\end{abstract}


\maketitle
The possibility of obtaining lasing into a scattering medium has been predicted in the sixties by Letokov \cite{Letokhov_NRA} and experimentally realized for bulk lasing material and scattering particles dispersed in liquid dye in the last decades\cite{Gouedard_RL_by_powder, Lawandy_Nature, Wiersma_Rew}. These pioneering experiments on random lasers (RL) have been characterized by a line narrowed frequency spectrum (from tens to some nanometers) on the top of the active medium fluorescence spectrum. This phenomenon, denoted as Intensity Feedback Random Lasers (IFRL), may be explained in the framework of diffusion approximation \cite{Wiermsa_Diff_Gain} in which all the properties of the photons propagation are defined by the transport mean free path $\ell$. This approach, which neglects the wavelike nature of light, is particulary useful in different scattering regimes where it may be used to predict particular properties of RL, like fluctuations \cite{ Lepri_Wiersma_Levi}, or the appearance of spikes in the spectrum at random positions\cite{ Mujumdar_Chaotic}. On the other hand, the diffusion approximation does not allow to predict the existence of RL peaks that have been more recently measured in numerous experiments \cite{Cao_RL_Action_Semiconductor}. These peaks, which appear at a fixed wavelength and whose light is emitted from localized portions of the disordered structure, may be associated with the presence of Anderson localized states \cite{Cao_spatial_confinement} or with highly scattering resonances dwelling in the disordered structures  \cite{Lagendijk_spatial_extent_RL}, that are commonly referred to as  Resonant Feedback Random Lasers (RFRL). This kind of lasing emission has been observed in very strong scattering systems (Zinc Oxide or Gallium Phosphide) with very small spatial extension, very small pump spot, or also with stripe shaped spot\cite{PhysRevB.59.15107}. In particular it has been shown\cite{PhysRevA.81.043830} that by increasing the pump spot of a RFRL a smooth spectrum is recovered.

\begin{figure}[!h]
\begin{center}
\includegraphics[width=8cm]{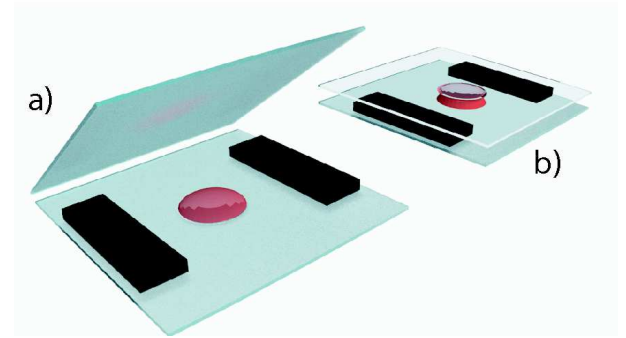} 
\caption{ (Color online) Sample preparation: In a) a drop of the solution containing scattering particles and dye is placed on a glass substrate with two plastic spacers. In b) a second substrate is placed on the top of the sample squeezing the drop.  }\label{Fig_Sample}
\end{center}
\end{figure}

Up to now different theoretical approaches have been suggested to account for the different aspects of RLs. The RL spectrum has been investigated through the strong nonlinear interaction of RL individual lossy modes \cite{Tureci02052008}, while an analogy with condensed matter physics a spin-glass-like model \cite{Glassy_Light_PRL} predicts the existence of various thermodynamic phases including a mode locked condition, in which lasing resonances are synchronized. A different approach valid in the condition in which the number of resonances pushes to infinity, allows to cast an equation for the spectral shape of the intensity predicting a behavior as a function of energy  in agreement with the experiments \cite{Conti_condensation}.

\begin{figure}[h!]
\begin{center}
\includegraphics[width=8cm]{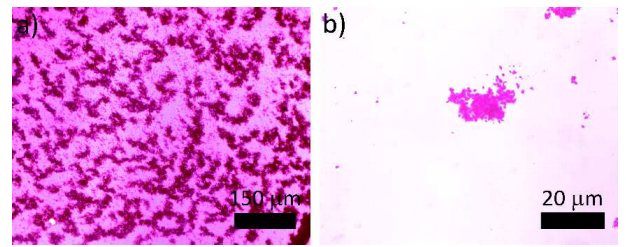} 
\caption{ (Color online) Sample before (a) and after (b) the isolation protocol.  }\label{Fig_Sample_micro}
\end{center}
\end{figure}

This review follows a recent letter \cite{Leonetti2011} in which it has been demonstrated that the particular regime (so far called either resonant or diffusive) in which RL may appear is driven by the degree of mode synchronization. This has been demonstrated by following an experimental approach consisting of several features:

I) A new random lasing system, in which a single strong scattering cluster is isolated and embedded in a dye solution.

II) A new pumping scheme in which the cluster is optically pumped by using stimulated emission generated in the dye thus allowing to obtain a \emph{"directional pumping"}

III) An engineered scheme to collect the RL spectrum.

Our approach allows to tune and monitor the amount of activated lasing modes in a single micron-sized titanium dioxide cluster, and to measure the inter mode correlation and the transition from an uncorrelated (with a RFRL-like spectrum) to a strongly correlated random lasing (with a IFRL-like spectrum).

In this paper we report extensive details on this experimental approach including the analysis of the random laser distribution of the intensity that was not previously reported.\begin{figure}[h!]
\includegraphics[width=\columnwidth]{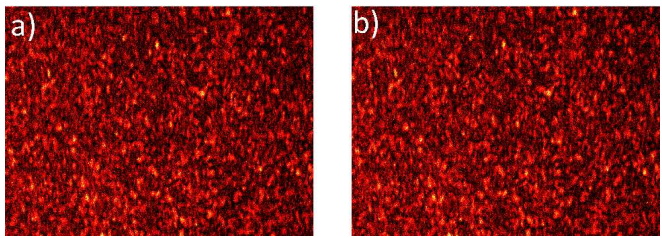}
\caption{(Color online) Passive speckle pattern from a single cluster obtained before (panel a) and after (panel b) 150 pump shots. The correlation between the two images is 0.85.\label{Spekle}}
\end{figure}

\section{Sample preparation}
\begin{figure}[]
\begin{center}
\includegraphics[width=8cm]{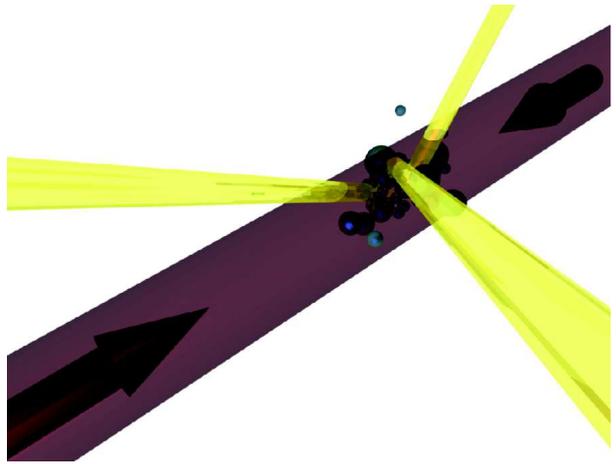} 
\caption{ (Color online) A schematic view of a single cluster placed in the center of   a stripe of a population inverted material. Stimulated emission arriving from the directions indicated by the two red arrows stimulates lasing modes}\label{Directional_pump}
\end{center}
\end{figure}

The majority of previous experiments have been performed by exploiting micron-size powder arranged in macroscopic (millimeter-size) samples. As it has been demonstrated in a recent paper, the extension of random lasing modes may be as small as some microns \cite{PhysRevLett.98.143901} thus if one wants to study the properties of a small set of modes the dimension of the system should be comparable. A possible strategy is the one used in the work by Fallert et al.\cite{Fallert_coexistence_nature} in which the spatial structure of random lasing modes has been studied in an compact cluster of zinc oxide powder. This material has sçeveral advantages: very high index contrast and very high damage threshold. Random lasing may be also achieved by using titanium dioxide dispersions in solutions containing Rhodamine \cite{Lawandy_Nature}; even if less efficient this system possesses a fundamental difference from Zinc Oxide: gain and scattering are hosted by two different materials and we exploit such feature to achieve control over the modes. To obtain a micron-size cluster structure embedded in the rhodamine solution we proceed as follows:
First we produce a solution containing 1 vol.\%  of titanium dioxide powder [ Titanium(IV) oxide, 89490 Sigma-Aldrich, particles size $\leq$ 1 m$\mu m$] in ethanol, then another solution containing 1\% volume of Rhoramine B (Sigma-Aldrich 242425) in ethylene-glycol. The first and the second solutions are mixed in equal part to obtain a third solution that is placed in an ultrasound bath for half an hour. A single drop (from two to five microliters) is placed on a thin microscopy coverslip (figure \ref{Fig_Sample}-a). After placing two plastic spacers (thickness $S =50$ $\mu m$) the drop is squeezed with a second coverslip (figure \ref{Fig_Sample}-b). In this thin system we observed on the bottom glass the formation of clusters of titanium dioxide with size of the order of tens of microns. Figure \ref{Fig_Sample_micro} a shows images obtained by an optical microscope in transmission geometry of the sample in which the clusters appear as black spots in a rhodamine (pink) bath.

\begin{figure}
\includegraphics[width=\columnwidth]{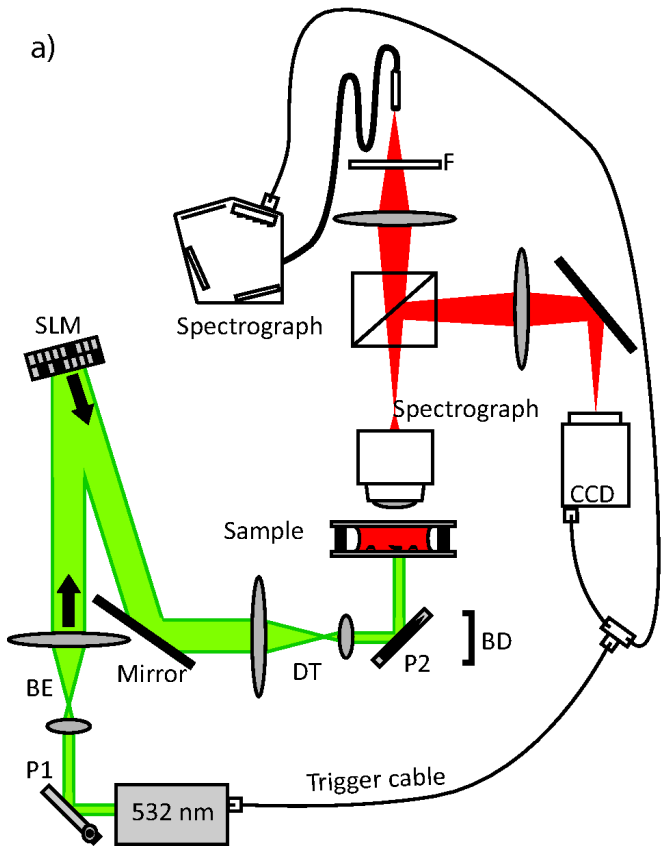}
\caption{ (Color online) A sketch of the experimental setup.  The vertically polarized laser beam is first reflected by a high power dichroic polarizing mirror  (P1) to clean residual unpolarized light and then passes through a 5$\times$ beam expander (BE) before impinging on the spatial light modulator (SLM). A demagnifying telescope (DT) images the spatial light modulator on the sample after a polarizing dichroic mirror (P2), which selects horizontal polarization, while residual vertically polarized light is absorbed by a beam dump (BD). Emitted  light is collected by a microscope objective (numerical aperture 0.40, working distance 6 mm), to be imaged by an enlarging telescope (x10) on a 50 $\mu$m core fiber, after being filtered by an edge filter(F) cutting residual green laser light. RL light is divided by a beam splitter to feed a spectrograph and a CCD imaging camera that are triggered by the pumping laser. \label{setup}}
\end{figure}

After some minutes since the sample fabrication the clusters appear to be completely formed and stable in their position and shape. Clusters possess random shape and size, and usually appear close together, a condition making difficult the experimental study of an isolated cluster. To obtain an isolated cluster we proceeded as follows: once individuated the target cluster, we pump by with a beam from a Nd:Yag pulsed laser ( fluence 0.02 $nJ/ \mu m^2$) by using the setup described below. The beam heats the liquid solution generating a flux that bring slowly the other clusters away from the cluster placed in the center of the pump spot (we see the flux acting principally at the edges of the pumped region). Figure \ref{Fig_Sample_micro}-b) shows a single cluster after this isolation protocol.

We observed by optical imaging that clusters positions remains fixed for days.  The RL analyzed below are critically dependent on cluster position, thus we further investigated the cluster stability by monitoring fluctuations in the speckle pattern obtained by shining a single cluster by a continuous Helium-Neon laser before and after 150 shots from the pump laser following reference \cite{Mujumdar_Chaotic}. The two speckle patterns in figure \ref{Spekle} show a correlation of about 0.85 denoting a static disorder.

\section{Experimental Setup}
\begin{figure}
\includegraphics[width=\columnwidth]{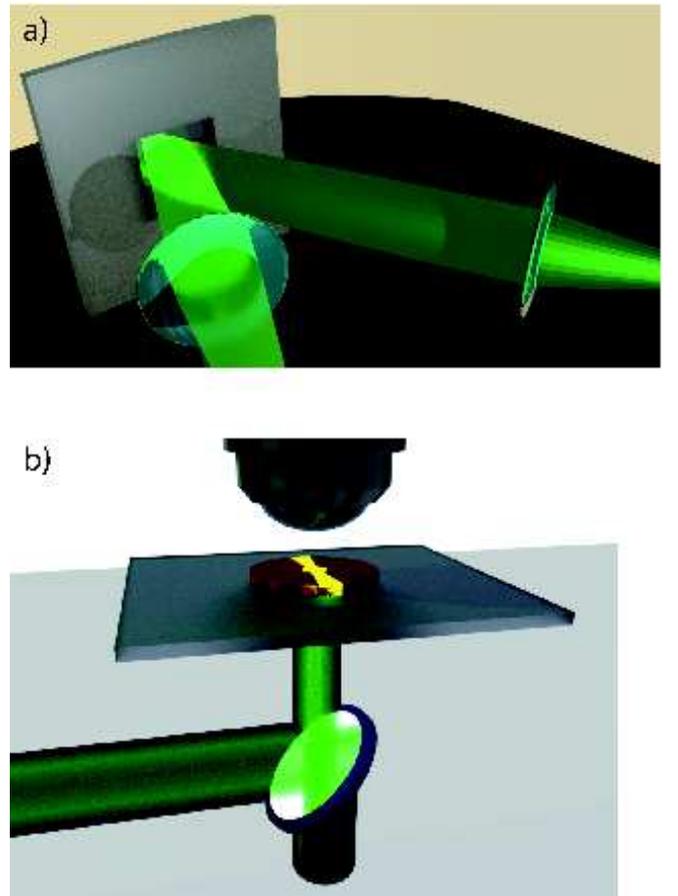}
\caption{(Color online) (a) shows an explicative perspective view of the SLM; (b) represents the sample under pumping. \label{setup_persp}}
\end{figure}
In this paper we will demonstrate that the distribution of the spectral peaks , the inter mode correlation, and the spatial distribution of the intensity may be controlled by the pumping shape, that directly affects the interaction between resonances. The modes interact due to the fact that the considered cavities are open, so that leaked radiation may travel between different modes of random the structure. This picture features a degree of interaction strongly affected by the distance between the modes, i.e. modes living close together will interact more strongly than modes that dwell at distant positions. In previous experiments peaks of the RFRL spectrum have been excited by a circularly focused beam. This represent a strong constraint in the study of RL properties: if all modes are activated in a small area they necessarily stay close together meaning their interaction is always strong. We developed a different approach that allows to tune the degree of interaction between the modes (we call it ``\emph{directional pumping}''). In a standard RL experiment, energy is brought to the modes by an external laser pump so that the absorption and re-emission process does not possess any specific direction. A directionality may be introduced by using a stripe shaped pump beam in which a flux of stimulated emission is generated along the stripe direction. A similar approach, that has been already exploited to study diffusive system \cite{Leonetti:11}, allows to turns on modes that are much less interacting than the circular pumping configuration, in fact modes does not have the constraint to lie close, instead they have only to comply with the condition to being efficiently coupled with the input direction.

A scheme of the setup is given in figure \ref{setup}. The pulse of the pump laser is generated by  a frequency doubled Nd:YAG pulsed laser, frequency 532 nm,  maximum pulse energy 125 mJ, pulse duration 9 ns. To achieve a complete control on the shape of the pump beam we used a reflective spatial light modulator ( figure \ref{setup_persp}-a) from Holoeye (LC-R 1080, 1920x1200 pixels) in amplitude modulation configuration \cite{Savage2009}. Light emitted from the sample is collected by a microscopy objective ( figure \ref{setup_persp}-b)) and sent to a spectrograph and to an imaging camera by a 50:50 beam splitter. The spectrum is collected by a 50 $\mu$m fiber placed in the image plane of an enlarging (10 $\times$) telescope in transmission geometry, thus collecting light from 5 $\mu$m area of at the center of the sample. A colored filter eliminates residual pump light. Spectra are obtained by a 303mm focal length spectrograph (Andor, Shamrock 303) connected to a low noise Charge Coupled Device array (Andor, iDus Spectroscopy CCD). Image of the sample is obtained by a standard scientific camera (Pixelink model PL-B776F).

\section{Random laser activated by directional pumping}
\begin{figure}
\includegraphics[width=\columnwidth]{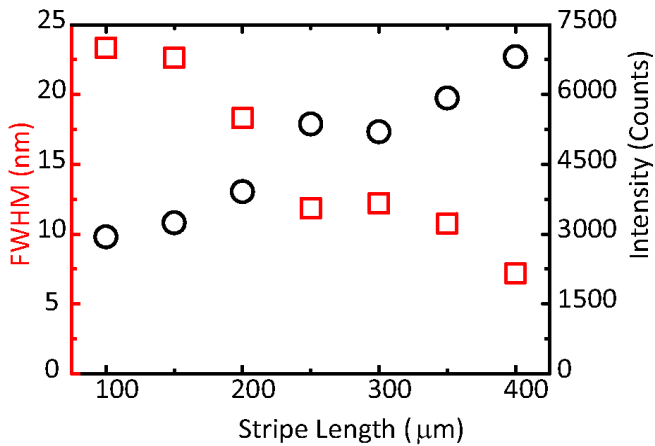}
\caption{ (Color online) FWHM (red squares) and peak intensity (black circles) as a function of the length of the illuminated stripe. The measurement has been performed with a pump power of 0.1 nJ/$\mu m^2$ pulse.   \label{Stripe_pump}}
\end{figure}
In this section we will describe the properties of a directionally pumped random laser. We will study a single cluster of titanium dioxide embedded in rhodamine solution. In this configuration the properties of the RL emission from the cluster not only depend on the energy that arrives ``\emph{directly}'' from the external pump to the cluster but also on the energy arriving from the emission generated in the neighborhood of the cluster. This is demonstrated  by monitoring the emission of a single cluster shone  by a beam shaped in the stripe configuration (16 $\mu$m thickness) when varying the length of the long side of the stripe that affects the intensity of the stimulated emission, which is  the "\emph{indirect}" pumping shining the cluster. In figure \ref{Stripe_pump} we shown intensity and full width at half maximum (FWHM) of the emission as a function of the length of the stripe shining the sample.

 \begin{figure}[h!]
\includegraphics[width=\columnwidth]{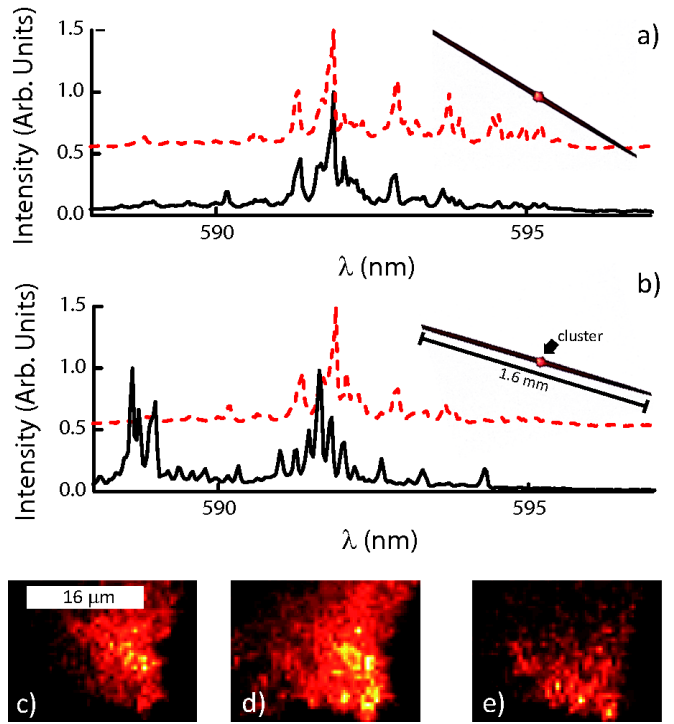}
\caption{(Color online) (a) Spectrum (continuous line) obtained with configuration 1 (see text) together with the spectrum obtained with the same orientation but with a 32 $\mu m$ thick stripe (configuration 2, dashed red line arbitrarily translated along the ordinate). (b)  Spectrum obtained by pumping with configuration 3 obtained by tilting the stripe by 15 degrees with respect to configuration 1. Spectrum obtained from configuration 1 reported as a dashed red line and arbitrarily translated along the ordinate for comparison. Panel c) d) and e) show the spatial distribution of intensity for configurations 1 , 2 and 3. The inset represent the pumping shape of configuration 1 and 3. Al measurements are obtained with a  0.1 nJ/$\mu m^2$ pump fluence\label{Stripe_properties}}
\end{figure}

The growth of the intensity measured when the stripe is longer may be only explained with the presence of an indirect and directional pumping. The narrowing of the FWHM  proves the onset of the lasing regimes at long values of the stripe length. We stress that our confocal configuration in which a fiber is placed in an enlarged image plane, allowing to spatially filter all the light that is not arising from the cluster, is fundamental to measure the described phenomena; otherwise the fluorescence from the whole pumped area would hide the laser emission from the cluster.

Then we study different stripe configurations all sharing the same lenght (1.6 mm) but differing in orientation and thickness. In figure \ref{Stripe_properties}a (black continuous line) we report the average spectrum on 100 acquisitions or ``\emph{shots}'' from a pumping stripe with thickness 16 $\mu m$ (orientation thickness and pump power on the stripe are conditions that we identify as ``configuration 1''). Noticeably, the presence of sharp (sub-nanometric) and repeatable peaks in the spectrum is connected to the appearance of bright spots  in the spatial distribution of the intensity (figure \ref{Stripe_properties} panel c), thus these features are due to modes dwelling in the cluster structure. Increasing the stripe thickness (32 $\mu m$, configuration 2) produces a larger number of activated modes  (figure \ref{Stripe_properties} a red dashed line) and new bright spots appear in the spatial distribution of intensity (figure \ref{Stripe_properties} panel d). The particular set of activated modes is nearly unaffected by the stripe thickness while is strongly sensitive to the stripe orientation. The spectrum in figure \ref{Stripe_properties}-b (continuous black line) has been obtained for a 16 $\mu m$ thick stripe tilted by 15$^\circ$ (configuration 3) with respect to the previous measurement (reported for visual comparison by the red dashed line in the same panel). Panel \ref{Stripe_properties}-e) shows the relative spatial distribution of the intensity, and that this 15$^\circ$ tilt also affects the location where hot-spots appear.

\begin{figure}
\includegraphics[width=\columnwidth]{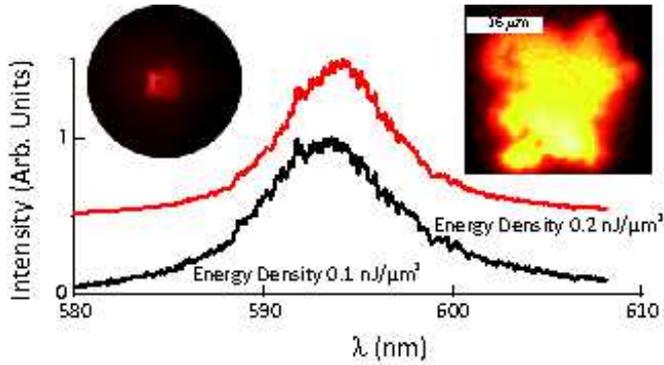}
\caption{ (Color online) Spectrum obtained for two different values of the pump power with a circular pump spot (left inset) on the  cluster in figure \ref{Stripe_properties}. An image of the intensity distribution of the cluster is shown in the right inset. \label{Stripe_circular_properties}}
\end{figure}

A different RL regime is obtained when directionality is absent, that is when using a circular pump spot. In figure \ref{Stripe_circular_properties} we show the spectrum obtained from the same cluster of figure \ref{Stripe_properties} (we identify this particular cluster as C1) with a circular pump  of  1 mm  diameter (a sketch is reported in the left inset of the figure) for two different values of the pump fluence. The retrieved spectrum in this case appears as a line narrowed smooth peak with tiny features repeatable from shot to shot measurement. The spatial distribution of the intensity (left inset) appears smooth and lacks the bright spots that can be seen in the stripe case.

\section{Transition from a spiky to a smooth random laser}
With the previously reported measurements we recognized the possibility to turn lasing emission within two possible emission regime. In the first case, by using a stripe shaped pump we retrieve a RFRL while a circular pump results in an IFRL emission. In this section we  demonstrate that it is possible, by controlling the pump shape, to turn gradually, from one regime to the other and control the degree of ``\emph{spikiness}'' of the spectrum.
\begin{figure}
\includegraphics[width=\columnwidth]{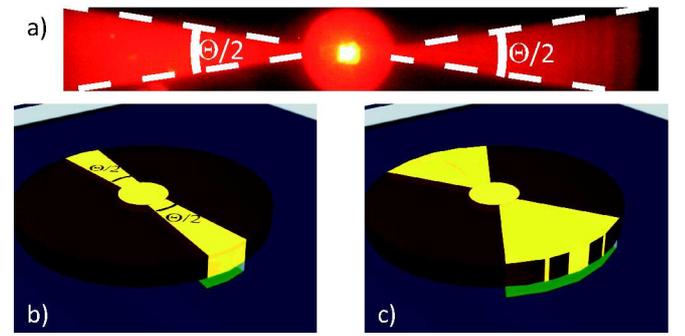}
\caption{ (Color online) a) An image of the sample shined with the "pie shaped pumping". In b) and c) A perspective scheme of the pumped sample for two different values of theta. \label{Pie_pump}}
\end{figure}

To this aim we engineered a more complex pumping geometry: the ``pie shaped pumping'' (figure \ref{Pie_pump}). The excited area consists of a disk (150 $\mu m$ in diameter) centered on the cluster (to assure homogeneous pumping even to the largest clusters) to which two symmetrical wedges of much larger radius ($1 mm$ in diameter) and controllable orientation and aperture angle ($\Theta/2$) are added, serving as launch pads for the directional stimulated emission. A single wedge configuration leads to similar results but proved to be hydrodynamically less stable. The central circle places the cluster barely below the lasing threshold preparing the system for lasing once the wedges are turned on. $\Theta$ controls the angular aperture with which stimulated emission is produced and therefore controls the number of modes expected to be excited.

\begin{figure}
\includegraphics[width=\columnwidth]{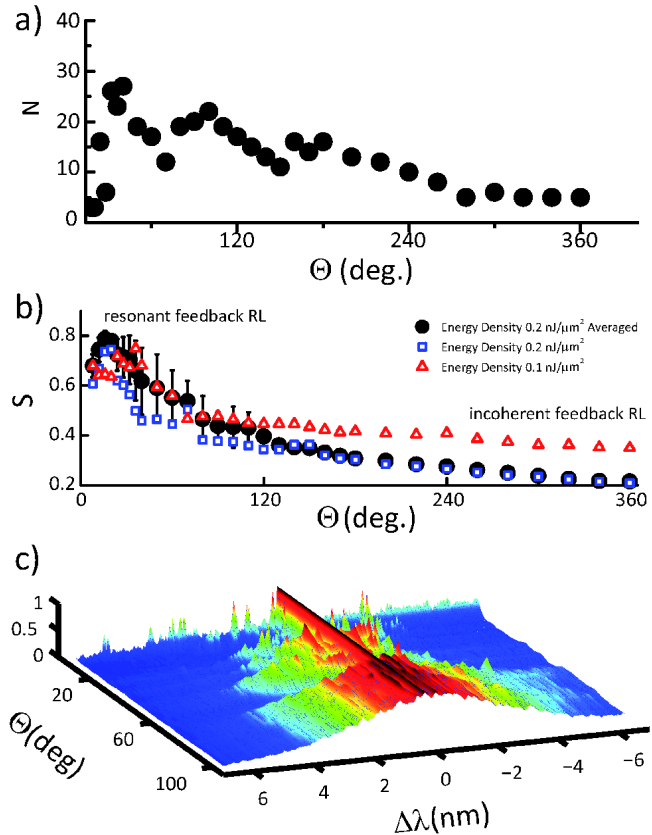}
\caption{(Color online) Number of modes $N$ (a) and spikiness (b) as a function of $\Theta$. Squares and triangles correspond to different pump energies for  the cluster C1. Circles (error bars) corresponds to the average (standard deviation) of 5 measurement from different clusters. The three dimensional graph on panel c) shows normalized spectra ( average over 100 shots with a fluence of 0.2 nJ/$\mu$m$^2$ )  for different $\Theta$. Spectra are arbitrarily shifted in frequency to superimpose intensity maxima.  $\Delta\lambda$ is the wavelength shift from the most intense peak. \label{Spikiness}}
\end{figure}

We note that the number of modes is strongly affected by the $\Theta$ parameter: in figure \ref{Spikiness}-a we plot the number of peaks in the spectrum $N$(that is the number of local maxima of the spectrum measured after a smoothing has been performed to eliminate noise effects) as a function of  $\Theta$. The number of modes in the spectrum first grows because of the increased width of the input directions allows for the excitation of more modes and then decreases because multiple peaks merge into a single narrowed  IFRL.
To classify a RL into the IFRL or RFRL categories we measure its \emph{spikiness} $S$ which is the amount of high frequency components of the Fourier Transform  (FTS) of the emission.  We define $S$ as the high-frequency fraction of the total FTS area, i.e., the area above a frequency threshold. As a cutoff we defined  $K$=1.20 nm$^{-1}$ in the horizontal scale of the FTS, then we calculate $S$ as the area of the FTS lying in the high period part from $K$ (i.e., corresponding to periods greater than $K$). $S$ returns a value close to one for very spiky spectra, while a value close to 0 is retrieved for smooth spectra.

Figure \ref{Spikiness}-b shows $S$ versus $\Theta$ for two pump energies (squares and triangles) for cluster  C1  and  averaged over 5 different clusters (circles). All curves display the same trend evidencing a transition: after a rapid growth corresponding to an increase in fluence and number of excited modes (appearing on a smooth fluorescence spectrum), $S$ reaches a maximum (RFRL regime) after which the spectrum becomes smoother as $\Theta$ grows until an IFRL-like emission is achieved. In panel \ref{Spikiness}-c we show the evolution of the average spectra from the spiky to the smooth case. Note that smoothing at high $\Theta$ is not due to averaging because sharp peaks are absent in single shot spectra too.

\section{Correlation of the RL fluctuations}
In our experiments both IFRL and RFRL show intensity fluctuations from shot to shot, while the modes (spectral) positions remain unmodified. Nevertheless the properties of the fluctuations of a RFRL strongly differ from that of an IFRL. In particular for a RFRL intensity fluctuations are uncorrelated (i.e. modes oscillate independently, that is, a mode can grow in intensity while the others decrease) while for a IFRL a decrease (increase) of the intensity at one frequency is strongly correlated to the trend of the intensity for all the other modes.

\begin{figure}
\includegraphics[width=\columnwidth]{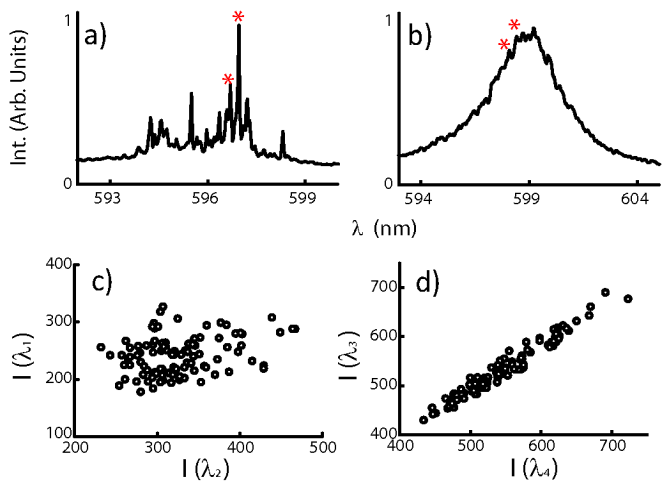}
\caption{(Color online) (a,b) show normalized average spectra from cluster C1 for $\Theta$=18$^\circ$ and $\Theta$=360$^\circ$ respectively. Asterisks indicate the position of the modes analyzed in panels  c and d. In panel (c) intensity values of the modes at wavelengths $\lambda_1$ =597.2 nm and $\lambda_2=$596.7 nm obtained for 100 single shot realization in the pumping configuration with $\Theta$=18$^\circ$. (d) as in (c) for  wavelengths $\lambda_3$=598.4 nm  and $\lambda_4$=598.7 nm and $\Theta$=360$^\circ$.  \label{correlation}}
\end{figure}

Let us consider the intensity of two modes of a RFRL (spectrum shown in figure \ref{correlation}-a obtained with $\Theta =18^\circ$) and of a IFRL (spectrum shown in figure \ref{correlation}-b) obtained with $\Theta =360^\circ$  from the same sample (cluster C1) with two different pumping. In panel \ref{correlation}-c we plot 100 values of the intensity obtained for 100 single shot measurements in the small $\Theta$ configuration. The correlation between the two intensity ensembles is calculated by using the Pearson correlation defined as follows:

\begin{equation}
C_p=\frac{\sum_{i=1}^{N}[I(\lambda_1)_i-\overline{I}(\lambda_1)][I(\lambda_2)_i-\overline{I}(\lambda_2)]}{\sqrt{\sum_{i=1}^{N}[I(\lambda_1)_i-\overline{I}(\lambda_1)]^2}\sqrt{\sum_{i=1}^{N}[I(\lambda_2)_i-\overline{I}(\lambda_2)]^2}}
\end{equation}

where $\overline{I}$ represent average over the ensemble. If two modes show a $C_p$ close to 0 their fluctuations are uncorrelated while correlated modes results in a $C_p$ close to 1.

\begin{figure}
\includegraphics[width=\columnwidth]{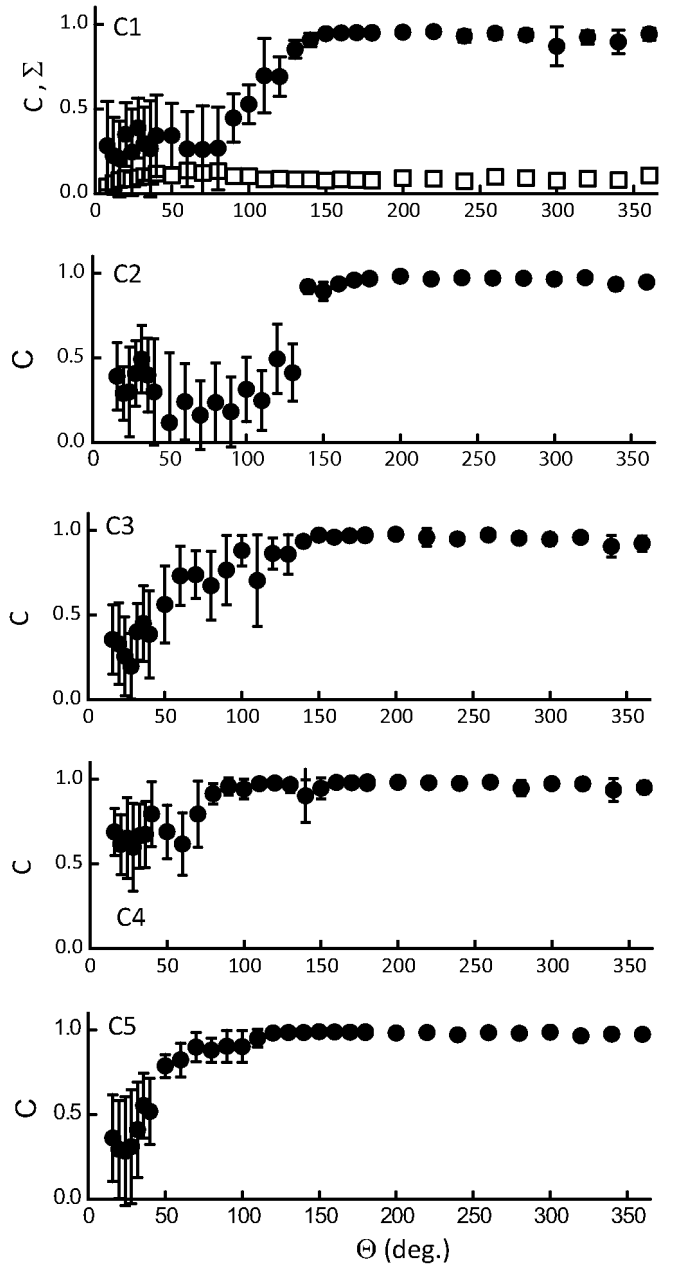}
\caption{Average correlation C as a function of $\Theta$ obtained for five different clusters. In the graph for C1 also $\Sigma$ is plotted as a function of $\Theta$ as open squares.\label{correlation_TOT}}
\end{figure}

The correlation for the modes of figure \ref{correlation}-a) is 0.47 while in for the modes in figure \ref{correlation}- b) correlation is 0.97.
Then we calculate the average correlation $C$ (figure \ref{correlation_TOT}) as the average of $C_p$ over 105 wavelength pairs, corresponding to the 15 more intense peaks, demonstrating that modes of an IFRL are, on average, much more correlated than that of an RFRL.

Figure \ref{correlation_TOT} shows the average correlation $C$  versus $\Theta$ for 5 different clusters named from C1 to C5. The onset of a strongly correlated regime is obtained for $\Theta\cong 120^\circ$. The fact that the same transition has been observed in all the considered samples reveals an universal trend in which $C$ $>0.8$ when $\Theta>$180$^\circ$.
The reported behavior of $C$  at large $\Theta$  is due to interaction and not to a lowering of the overall fluctuations. This is demonstrated in the graph for C1 which shows $C$ together with relative fluctuation $\Sigma$ defined as $\sigma(\lambda)/\overline{I(\lambda)}$ [where $\sigma(\lambda)$ is the standard deviation of the intensity for the frequency $\lambda$ over 100 shots] averaged for the 15 more intense modes of cluster C1. The fact that $\Sigma$ is nearly constant for all the values of $\Theta$ investigated allows to exclude artifacts.

\section{Measurement at fixed volume}

\begin{figure}
\includegraphics[width=\columnwidth]{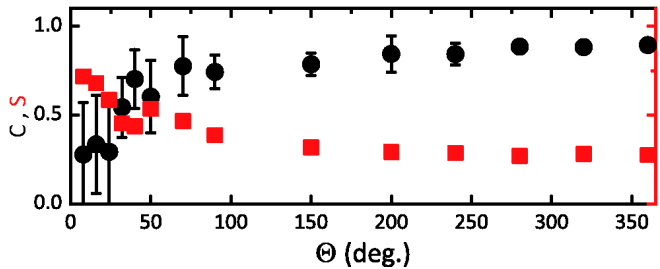}
\caption{ (Color online) $C$ (circles) and $S$ (squares)  as a function of $\Theta$ in the fixed volume configuration. Pumping is 0.1 nJ/$\mu$m$^2$ \label{fixedvolume}}
\end{figure}

The parameter that controls the transition between uncorrelated and correlated random lasing is the angle $\Theta$ that defines the possible input directions on the cluster. To increase $\Theta$ we add to the inverted area angular sections of a disk. Although this choice clearly defines the input directions, a variation of $\Theta$ does not conserve volume of the illuminated part of the sample and total energy impinging on the sample. To confirm that the change in regime  is due the quality of the inter-mode interaction, we performed a measurement at fixed volume and energy. The measurement is very similar to the previous one with a central disk plus the angular sectors of a larger circumference, but in this case for every value of $\Theta$ we also adjusted the value of the radius $R$ of the external circumference in order to conserve the area of the illumination surfaces. In practice an increment of $\Theta$ corresponds in a lowering of the value of $R$ to preserve a fixed total laser spot of 0.06 mm$^2$.
Figure \ref{fixedvolume} shows the relative measurement of $C$ and $S$ as a function of $\Theta$  for this fixed volume configuration. This measurement allows to rule out the effects of volume or energy dependent terms, like amplified spontaneous emission.

It is also worth mentioning the fact that  $S$ and $C$ do not depend on how many wedges are used to compose the pumping area. Although we show results for only two wedges any number can be used with similar results.


\section{Spatial properties of the spatial distribution of intensity}

\begin{figure}
\includegraphics[width=\columnwidth]{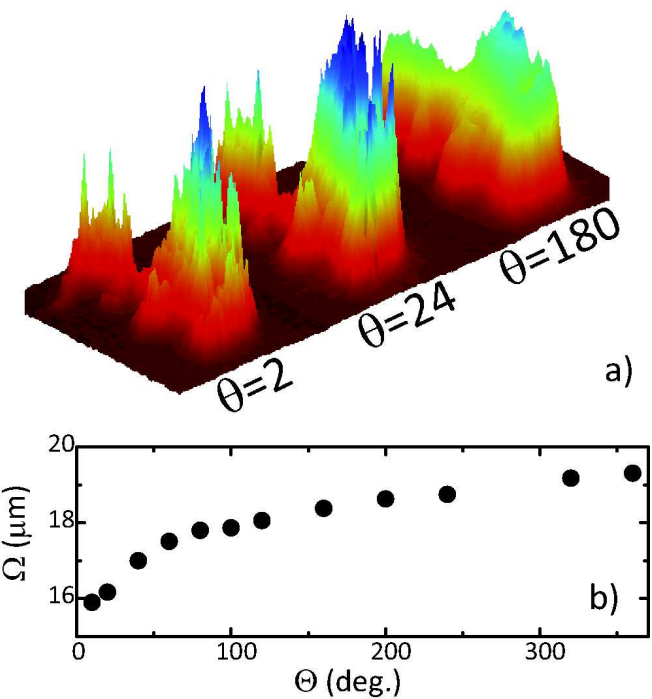}
\caption{ (Color online) (a) Surface plot of the intensity pattern for the same cluster for three different values of $\Theta$. (b)  The $\Omega$ is plotted as a function of the  parameter $\Theta$.  \label{IPS}}
\end{figure}

It has been shown in a previous paragraph (figures \ref{Stripe_properties} and \ref{Stripe_circular_properties}) that the spatial distribution of the intensity is strongly affected by $\Theta$: it is characterized by a profile with sharp features (hot spots) in the RFRL case while it displays  a smooth spatial profile for large $\Theta$ (see panel \ref{IPS} a). Here we investigate this feature  by studying the inverse participation ratio $P$ of the intensity distribution that quantifies the degree of confinement of the light in the sample \cite{Gentilini_09, Schwartz2007}. It may be calculated as
\begin{equation}
P=\frac{\int I^2 (x,y) dx dy}{\left[\int I(x,y)dx dy\right]^2 }
\end{equation}
where $I(x,y)$ is the measured spatial intensity distribution of the light. From $P$, that has the units of an inverse area one may obtain the extension $\Omega=\langle P^\frac{-1}{2}\rangle$  which has the dimension of length. In figure \ref{IPS}-b; $\Omega$ is plotted as a function of the  parameter $\Theta$. The the presence of sharp features for an intensity spatial distribution organized in a more localized fashion is signature of the fact that energy is stored in particular points of the sample and may not travel for one place to the other for low $\Theta$, coherently with the low values of $C$ measured for $\Theta< 100 ^\circ$. Conversely at high $\Theta$, a larger number of modes is activated, consequently all the cluster is filled with activated resonances and energy may flow trough the cluster resulting in a smooth spatial profile and a more correlated emission consistently  with an increased inter-mode interaction.

\section{Time resolved measurements}
The reported  behavior of $C$ and $S$ may be also obtain by a picosecond pump pulse (we used a 30 ps pulse frequency doubled Nd:Yag laser with 60 mJ/pulse maximum energy) that also allows to resolve temporal behavior of emitted light \cite{Siddique:96}.

\begin{figure}[h!]
\includegraphics[width=\columnwidth]{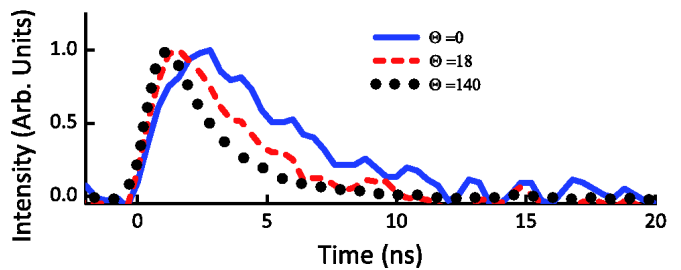}
\caption{ (Color online) Intensity as a function of time for the cluster C1 pumped with wedges of different aperture. Measurements are performed at a pump fluence of 0.1 nJ/$\mu$m$^2$)\label{fig:Timeresolved}}.
\end{figure}

Light is collected from a single cluster by a photodiode (1 nanosecond risetime and 1 ns falltime)  and a 300MHz oscilloscope. In figure \ref{fig:Timeresolved} we show results for cluster C1. Lasing in the correlated regime ($\Theta$=140 $^\circ$) shows shorter pulses with respect to the uncorrelated ($\Theta$=18 $^\circ$) random laser for all the 5 clusters. We obtained an estimate of the RL pulse duration by measuring the full width at half maximum of the peak in intensity of the average of 8 single shots time domain measurements retrieving an average shortening of about 30\% for the IFRL with respect to the RFRL. In the same figure \ref{fig:Timeresolved} we also report temporal shape of fluorescence obtained pumping the sample only with the central disk ($\Theta$=0 $^\circ$ ).

In the ordered case, pulse duration scales as 1/N \cite{YarivBook}, conversely in the disordered case is not at all expected that the pulse duration decreases substantially with the number of modes; on the other hand, it is reasonable that some shortening happens because the modes act synchronously, e.g., they tend to turn at the same time, because as one of them gets energy, it also feeds the others.

\section{Conclusions}
By exploiting directional pumping we are able, to control the activated modes of an RL and the onset by different random lasing regimes. The number and the set of activated modes is controlled by $\Theta$, a geometrical parameter that defines  the shape of the pumped area and the directions from which stimulated emission may reach the cluster. We demonstrated that when $\Theta$ grows, the RL undergoes a transition switching its emission from a  spiky (RFRL) to a smooth (IFRL) one. All the phenomena documented here may be accounted for in the framework of a mode locking transition that allows to explain spectral and temporal features and the onset of a correlated RL regime.

%


\textit{Acknowledgements} The work was supported by: ERC GRANT (FP7/2007-2013) n.201766;  CINECA; EU FP7 NoE Nanophotonics4Enery Grant No 248855; the Spanish MICINN CSD2007-0046 (Nanolight.es); MAT2009-07841 (GLUSFA) and Comunidad de Madrid S2009/MAT-1756 (PHAMA).

%
%

\begin{thebibliography}{24}
\expandafter\ifx\csname natexlab\endcsname\relax\def\natexlab#1{#1}\fi
\expandafter\ifx\csname bibnamefont\endcsname\relax
  \def\bibnamefont#1{#1}\fi
\expandafter\ifx\csname bibfnamefont\endcsname\relax
  \def\bibfnamefont#1{#1}\fi
\expandafter\ifx\csname citenamefont\endcsname\relax
  \def\citenamefont#1{#1}\fi
\expandafter\ifx\csname url\endcsname\relax
  \def\url#1{\texttt{#1}}\fi
\expandafter\ifx\csname urlprefix\endcsname\relax\def\urlprefix{URL }\fi
\providecommand{\bibinfo}[2]{#2}
\providecommand{\eprint}[2][]{\url{#2}}

\bibitem[{\citenamefont{Leonetti et~al.}(2011)\citenamefont{Leonetti, Conti,
  and Lopez}}]{Leonetti2011}
\bibinfo{author}{\bibfnamefont{M.}~\bibnamefont{Leonetti}},
  \bibinfo{author}{\bibfnamefont{C.}~\bibnamefont{Conti}}, \bibnamefont{and}
  \bibinfo{author}{\bibfnamefont{C.}~\bibnamefont{Lopez}},
  \bibinfo{journal}{Nat Photon} \textbf{\bibinfo{volume}{5}},
  \bibinfo{pages}{615} (\bibinfo{year}{2011}).

\bibitem[{\citenamefont{Letokhov}(1967)}]{Letokhov_NRA}
\bibinfo{author}{\bibfnamefont{V.}~\bibnamefont{Letokhov}},
  \bibinfo{journal}{Zh. Eksp. Teor. Fiz.} \textbf{\bibinfo{volume}{53}},
  \bibinfo{pages}{1442} (\bibinfo{year}{1967}).

\bibitem[{\citenamefont{Gouedard et~al.}(1993)\citenamefont{Gouedard, Husson,
  Sauteret, Auzel, and Migus}}]{Gouedard_RL_by_powder}
\bibinfo{author}{\bibfnamefont{C.}~\bibnamefont{Gouedard}},
  \bibinfo{author}{\bibfnamefont{D.}~\bibnamefont{Husson}},
  \bibinfo{author}{\bibfnamefont{C.}~\bibnamefont{Sauteret}},
  \bibinfo{author}{\bibfnamefont{F.}~\bibnamefont{Auzel}}, \bibnamefont{and}
  \bibinfo{author}{\bibfnamefont{A.}~\bibnamefont{Migus}}, \bibinfo{journal}{J.
  Opt. Soc. Am. B} \textbf{\bibinfo{volume}{10}}, \bibinfo{pages}{2358}
  (\bibinfo{year}{1993}).

\bibitem[{\citenamefont{Lawandy et~al.}(1994)\citenamefont{Lawandy,
  Balachandran, Gomes, and Sauvain}}]{Lawandy_Nature}
\bibinfo{author}{\bibfnamefont{N.~M.} \bibnamefont{Lawandy}},
  \bibinfo{author}{\bibfnamefont{R.~M.} \bibnamefont{Balachandran}},
  \bibinfo{author}{\bibfnamefont{A.~S.~L.} \bibnamefont{Gomes}},
  \bibnamefont{and} \bibinfo{author}{\bibfnamefont{E.}~\bibnamefont{Sauvain}},
  \bibinfo{journal}{Nature} \textbf{\bibinfo{volume}{368}},
  \bibinfo{pages}{436} (\bibinfo{year}{1994}).

\bibitem[{\citenamefont{Wiersma}(2008)}]{Wiersma_Rew}
\bibinfo{author}{\bibfnamefont{D.~S.} \bibnamefont{Wiersma}},
  \bibinfo{journal}{Nat Phys} \textbf{\bibinfo{volume}{4}},
  \bibinfo{pages}{359} (\bibinfo{year}{2008}).

\bibitem[{\citenamefont{Wiersma and Lagendijk}(1996)}]{Wiermsa_Diff_Gain}
\bibinfo{author}{\bibfnamefont{D.~S.} \bibnamefont{Wiersma}} \bibnamefont{and}
  \bibinfo{author}{\bibfnamefont{A.}~\bibnamefont{Lagendijk}},
  \bibinfo{journal}{Phys. Rev. E} \textbf{\bibinfo{volume}{54}},
  \bibinfo{pages}{4256} (\bibinfo{year}{1996}).

\bibitem[{\citenamefont{Lepri et~al.}(2007)\citenamefont{Lepri, Cavalieri,
  Oppo, and Wiersma}}]{Lepri_Wiersma_Levi}
\bibinfo{author}{\bibfnamefont{S.}~\bibnamefont{Lepri}},
  \bibinfo{author}{\bibfnamefont{S.}~\bibnamefont{Cavalieri}},
  \bibinfo{author}{\bibfnamefont{G.-L.} \bibnamefont{Oppo}}, \bibnamefont{and}
  \bibinfo{author}{\bibfnamefont{D.~S.} \bibnamefont{Wiersma}},
  \bibinfo{journal}{Phys. Rev. A} \textbf{\bibinfo{volume}{75}},
  \bibinfo{pages}{063820} (\bibinfo{year}{2007}).

\bibitem[{\citenamefont{Mujumdar et~al.}(2007)\citenamefont{Mujumdar,
  T{\"u}rck, Torre, and Wiersma}}]{Mujumdar_Chaotic}
\bibinfo{author}{\bibfnamefont{S.}~\bibnamefont{Mujumdar}},
  \bibinfo{author}{\bibfnamefont{V.}~\bibnamefont{T{\"u}rck}},
  \bibinfo{author}{\bibfnamefont{R.}~\bibnamefont{Torre}}, \bibnamefont{and}
  \bibinfo{author}{\bibfnamefont{D.~S.} \bibnamefont{Wiersma}},
  \bibinfo{journal}{Phys. Rev. A} \textbf{\bibinfo{volume}{76}},
  \bibinfo{pages}{033807} (\bibinfo{year}{2007}).

\bibitem[{\citenamefont{Cao et~al.}(1999{\natexlab{a}})\citenamefont{Cao, Zhao,
  Ho, Seelig, Wang, and Chang}}]{Cao_RL_Action_Semiconductor}
\bibinfo{author}{\bibfnamefont{H.}~\bibnamefont{Cao}},
  \bibinfo{author}{\bibfnamefont{Y.~G.} \bibnamefont{Zhao}},
  \bibinfo{author}{\bibfnamefont{S.~T.} \bibnamefont{Ho}},
  \bibinfo{author}{\bibfnamefont{E.~W.} \bibnamefont{Seelig}},
  \bibinfo{author}{\bibfnamefont{Q.~H.} \bibnamefont{Wang}}, \bibnamefont{and}
  \bibinfo{author}{\bibfnamefont{R.~P.~H.} \bibnamefont{Chang}},
  \bibinfo{journal}{Phys. Rev. Lett.} \textbf{\bibinfo{volume}{82}},
  \bibinfo{pages}{2278} (\bibinfo{year}{1999}{\natexlab{a}}).

\bibitem[{\citenamefont{Cao et~al.}(2000)\citenamefont{Cao, Xu, Zhang, Chang,
  Ho, Seelig, Liu, and Chang}}]{Cao_spatial_confinement}
\bibinfo{author}{\bibfnamefont{H.}~\bibnamefont{Cao}},
  \bibinfo{author}{\bibfnamefont{J.~Y.} \bibnamefont{Xu}},
  \bibinfo{author}{\bibfnamefont{D.~Z.} \bibnamefont{Zhang}},
  \bibinfo{author}{\bibfnamefont{S.-H.} \bibnamefont{Chang}},
  \bibinfo{author}{\bibfnamefont{S.~T.} \bibnamefont{Ho}},
  \bibinfo{author}{\bibfnamefont{E.~W.} \bibnamefont{Seelig}},
  \bibinfo{author}{\bibfnamefont{X.}~\bibnamefont{Liu}}, \bibnamefont{and}
  \bibinfo{author}{\bibfnamefont{R.~P.~H.} \bibnamefont{Chang}},
  \bibinfo{journal}{Phys. Rev. Lett.} \textbf{\bibinfo{volume}{84}},
  \bibinfo{pages}{5584} (\bibinfo{year}{2000}).

\bibitem[{\citenamefont{{van der Molen}
  et~al.}(2007{\natexlab{a}})\citenamefont{{van der Molen}, Tjerkstra, Mosk,
  and Lagendijk}}]{Lagendijk_spatial_extent_RL}
\bibinfo{author}{\bibfnamefont{K.~L.} \bibnamefont{{van der Molen}}},
  \bibinfo{author}{\bibfnamefont{R.~W.} \bibnamefont{Tjerkstra}},
  \bibinfo{author}{\bibfnamefont{A.~P.} \bibnamefont{Mosk}}, \bibnamefont{and}
  \bibinfo{author}{\bibfnamefont{A.}~\bibnamefont{Lagendijk}},
  \bibinfo{journal}{Phys. Rev. Lett.} \textbf{\bibinfo{volume}{98}},
  \bibinfo{pages}{143901} (\bibinfo{year}{2007}{\natexlab{a}}).

\bibitem[{\citenamefont{Cao et~al.}(1999{\natexlab{b}})\citenamefont{Cao, Zhao,
  Ong, and Chang}}]{PhysRevB.59.15107}
\bibinfo{author}{\bibfnamefont{H.}~\bibnamefont{Cao}},
  \bibinfo{author}{\bibfnamefont{Y.~G.} \bibnamefont{Zhao}},
  \bibinfo{author}{\bibfnamefont{H.~C.} \bibnamefont{Ong}}, \bibnamefont{and}
  \bibinfo{author}{\bibfnamefont{R.~P.~H.} \bibnamefont{Chang}},
  \bibinfo{journal}{Phys. Rev. B} \textbf{\bibinfo{volume}{59}},
  \bibinfo{pages}{15107} (\bibinfo{year}{1999}{\natexlab{b}}).

\bibitem[{\citenamefont{El-Dardiry et~al.}(2010)\citenamefont{El-Dardiry, Mosk,
  Muskens, and Lagendijk}}]{PhysRevA.81.043830}
\bibinfo{author}{\bibfnamefont{R.~G.~S.} \bibnamefont{El-Dardiry}},
  \bibinfo{author}{\bibfnamefont{A.~P.} \bibnamefont{Mosk}},
  \bibinfo{author}{\bibfnamefont{O.~L.} \bibnamefont{Muskens}},
  \bibnamefont{and}
  \bibinfo{author}{\bibfnamefont{A.}~\bibnamefont{Lagendijk}},
  \bibinfo{journal}{Phys. Rev. A} \textbf{\bibinfo{volume}{81}},
  \bibinfo{pages}{043830} (\bibinfo{year}{2010}).

\bibitem[{\citenamefont{Tureci et~al.}(2008)\citenamefont{Tureci, Ge, Rotter,
  and Stone}}]{Tureci02052008}
\bibinfo{author}{\bibfnamefont{H.~E.} \bibnamefont{Tureci}},
  \bibinfo{author}{\bibfnamefont{L.}~\bibnamefont{Ge}},
  \bibinfo{author}{\bibfnamefont{S.}~\bibnamefont{Rotter}}, \bibnamefont{and}
  \bibinfo{author}{\bibfnamefont{A.~D.} \bibnamefont{Stone}},
  \bibinfo{journal}{Science} \textbf{\bibinfo{volume}{320}},
  \bibinfo{pages}{643} (\bibinfo{year}{2008}).

\bibitem[{\citenamefont{Angelani et~al.}(2006)\citenamefont{Angelani, Conti,
  Ruocco, and Zamponi}}]{Glassy_Light_PRL}
\bibinfo{author}{\bibfnamefont{L.}~\bibnamefont{Angelani}},
  \bibinfo{author}{\bibfnamefont{C.}~\bibnamefont{Conti}},
  \bibinfo{author}{\bibfnamefont{G.}~\bibnamefont{Ruocco}}, \bibnamefont{and}
  \bibinfo{author}{\bibfnamefont{F.}~\bibnamefont{Zamponi}},
  \bibinfo{journal}{Phys. Rev. Lett.} \textbf{\bibinfo{volume}{96}},
  \bibinfo{pages}{065702} (\bibinfo{year}{2006}).

\bibitem[{\citenamefont{Conti et~al.}(2008)\citenamefont{Conti, Leonetti,
  Fratalocchi, Angelani, and Ruocco}}]{Conti_condensation}
\bibinfo{author}{\bibfnamefont{C.}~\bibnamefont{Conti}},
  \bibinfo{author}{\bibfnamefont{M.}~\bibnamefont{Leonetti}},
  \bibinfo{author}{\bibfnamefont{A.}~\bibnamefont{Fratalocchi}},
  \bibinfo{author}{\bibfnamefont{L.}~\bibnamefont{Angelani}}, \bibnamefont{and}
  \bibinfo{author}{\bibfnamefont{G.}~\bibnamefont{Ruocco}},
  \bibinfo{journal}{Phys. Rev. Lett.} \textbf{\bibinfo{volume}{101}},
  \bibinfo{pages}{143901} (\bibinfo{year}{2008}).

\bibitem[{\citenamefont{{van der Molen}
  et~al.}(2007{\natexlab{b}})\citenamefont{{van der Molen}, Tjerkstra, Mosk,
  and Lagendijk}}]{PhysRevLett.98.143901}
\bibinfo{author}{\bibfnamefont{K.~L.} \bibnamefont{{van der Molen}}},
  \bibinfo{author}{\bibfnamefont{R.~W.} \bibnamefont{Tjerkstra}},
  \bibinfo{author}{\bibfnamefont{A.~P.} \bibnamefont{Mosk}}, \bibnamefont{and}
  \bibinfo{author}{\bibfnamefont{A.}~\bibnamefont{Lagendijk}},
  \bibinfo{journal}{Phys. Rev. Lett.} \textbf{\bibinfo{volume}{98}},
  \bibinfo{pages}{143901} (\bibinfo{year}{2007}{\natexlab{b}}).

\bibitem[{\citenamefont{Fallert et~al.}(2009)\citenamefont{Fallert, Dietz,
  Sartor, Schneider, Klingshirn, and Kalt}}]{Fallert_coexistence_nature}
\bibinfo{author}{\bibfnamefont{J.}~\bibnamefont{Fallert}},
  \bibinfo{author}{\bibfnamefont{R.~J.~B.} \bibnamefont{Dietz}},
  \bibinfo{author}{\bibfnamefont{J.}~\bibnamefont{Sartor}},
  \bibinfo{author}{\bibfnamefont{D.}~\bibnamefont{Schneider}},
  \bibinfo{author}{\bibfnamefont{C.}~\bibnamefont{Klingshirn}},
  \bibnamefont{and} \bibinfo{author}{\bibfnamefont{H.}~\bibnamefont{Kalt}},
  \bibinfo{journal}{Nat Photon} \textbf{\bibinfo{volume}{3}},
  \bibinfo{pages}{279} (\bibinfo{year}{2009}), ISSN \bibinfo{issn}{1749-4885}.

\bibitem[{\citenamefont{Leonetti and L\'{o}pez}(2011)}]{Leonetti:11}
\bibinfo{author}{\bibfnamefont{M.}~\bibnamefont{Leonetti}} \bibnamefont{and}
  \bibinfo{author}{\bibfnamefont{C.}~\bibnamefont{L\'{o}pez}},
  \bibinfo{journal}{Opt. Lett.} \textbf{\bibinfo{volume}{36}},
  \bibinfo{pages}{2824} (\bibinfo{year}{2011}).

\bibitem[{\citenamefont{Savage}(2009)}]{Savage2009}
\bibinfo{author}{\bibfnamefont{N.}~\bibnamefont{Savage}}, \bibinfo{journal}{Nat
  Photon} \textbf{\bibinfo{volume}{3}}, \bibinfo{pages}{170}
  (\bibinfo{year}{2009}).

\bibitem[{\citenamefont{Gentilini et~al.}(2009)\citenamefont{Gentilini,
  Fratalocchi, Angelani, Ruocco, and Conti}}]{Gentilini_09}
\bibinfo{author}{\bibfnamefont{S.}~\bibnamefont{Gentilini}},
  \bibinfo{author}{\bibfnamefont{A.}~\bibnamefont{Fratalocchi}},
  \bibinfo{author}{\bibfnamefont{L.}~\bibnamefont{Angelani}},
  \bibinfo{author}{\bibfnamefont{G.}~\bibnamefont{Ruocco}}, \bibnamefont{and}
  \bibinfo{author}{\bibfnamefont{C.}~\bibnamefont{Conti}},
  \bibinfo{journal}{Opt. Lett.} \textbf{\bibinfo{volume}{34}},
  \bibinfo{pages}{130} (\bibinfo{year}{2009}).

\bibitem[{\citenamefont{Schwartz et~al.}(2007)\citenamefont{Schwartz, Bartal,
  Fishman, and Segev}}]{Schwartz2007}
\bibinfo{author}{\bibfnamefont{T.}~\bibnamefont{Schwartz}},
  \bibinfo{author}{\bibfnamefont{G.}~\bibnamefont{Bartal}},
  \bibinfo{author}{\bibfnamefont{S.}~\bibnamefont{Fishman}}, \bibnamefont{and}
  \bibinfo{author}{\bibfnamefont{M.}~\bibnamefont{Segev}},
  \bibinfo{journal}{Nature} \textbf{\bibinfo{volume}{446}}, \bibinfo{pages}{52}
  (\bibinfo{year}{2007}), ISSN \bibinfo{issn}{0028-0836}.

\bibitem[{\citenamefont{Siddique et~al.}(1996)\citenamefont{Siddique, Alfano,
  Berger, Kempe, and Genack}}]{Siddique:96}
\bibinfo{author}{\bibfnamefont{M.}~\bibnamefont{Siddique}},
  \bibinfo{author}{\bibfnamefont{R.~R.} \bibnamefont{Alfano}},
  \bibinfo{author}{\bibfnamefont{G.~A.} \bibnamefont{Berger}},
  \bibinfo{author}{\bibfnamefont{M.}~\bibnamefont{Kempe}}, \bibnamefont{and}
  \bibinfo{author}{\bibfnamefont{A.~Z.} \bibnamefont{Genack}},
  \bibinfo{journal}{Opt. Lett.} \textbf{\bibinfo{volume}{21}},
  \bibinfo{pages}{450} (\bibinfo{year}{1996}).

\bibitem[{\citenamefont{Yariv}(1991)}]{YarivBook}
\bibinfo{author}{\bibfnamefont{A.}~\bibnamefont{Yariv}},
  \emph{\bibinfo{title}{Quantum Electronics}} (\bibinfo{publisher}{Saunders
  College, San Diego}, \bibinfo{year}{1991}).

\end{thebibliography}
%

\end{document}